# Germany's Tax Revenue and its Total Administrative Cost

**Christopher Mantzaris**
*University of Primorska, Slovenia;* 68223065{at}student.upr.si

and

**Ajda Fošner**
*University of Primorska, Slovenia*

Abstract

Tax administrative cost reduction is an economically and socially desirable goal for public policy. This article proposes total administrative cost as percentage of total tax revenue as a vivid measurand, also useful for cross-jurisdiction comparisons. Statistical data, surveys and a novel approach demonstrate: Germany's 2021 tax administrative costs likely exceeded 20% of total tax revenue, indicating need for improvement of Germany's taxation system – and for the many jurisdictions with similar tax regimes. In addition, this article outlines possible reasons for and implications of the seemingly high tax administrative burden as well as solutions.

**Introduction**

**Background**

A priori, it is desirable to reach goals as efficiently, meaning with spending as little resources –or 'cost'–, as possible. To minimise waste of society's resources, the aim should therefore also be, to reach the goals of taxation with as little cost as possible. The 'hole in the redistribution bucket' can be a suitable imagery. This article examines those costs, which directly arise with tax revenue, and not how well the current situation reaches the goals of taxation.

This article defines **tax administration cost** as costs –whether in currency, labour or other forms– for the *revenue side* of taxes; for example for preparing documents, declaring or auditing. Those costs arise on the government side. There, it includes tax auditors, collectors, and so forth. Costs for legislation and policy creation may not clearly arise for tax *revenue* or at least are not easily distinguishable. Those should perhaps more fittingly be in another category and have the name 'legislative tax costs' or the like. The bulk of costs, however, arises on the taxpayer side. Because most jurisdictions transfer the majority of tax administrative burden onto the taxpayers, who need to prepare, declare etc taxes themselves, which often requires extensive time and not seldomly expensive professional assistance. The government then only roughly or randomly audits a fraction of those tax declarations, in addition to its other roles such as other areas of tax enforcement and collection. The taxpayer's cost can be further subdivided, see under Classification.

Notably, tax administrative costs are not the only costs for society associated with taxation. The other two include costs for the spending part of taxes – such as for legislative or executive politicians, who debate how much, for what and where to spend; or officials and company employees, who need to work for example on tendering or procurement. The third group of 'welfare' or 'deadweight loss' is the inefficiency of the allocation of capital by the government: Logic and historical experiences indicate profit-orientation to be the most efficient system available when it comes to resource allocation for a society's wealth creation (Smith, 1776, p. 90; Ricardo, 1912, p. 48; Anderton, 2015, pp. 23, 165, and 442). The government taking resources from profit-oriented individuals and companies for 'non-profit' or 'loss-bearing' endeavours may therefore also constitute (opportunity) costs.

This research focuses on the first-mentioned group because it is

1. the most objectively measurable of the three,
2. the best visualisation of 'transaction cost' of money from citizens to government, and
3. most useful to know for society and public policy: It is likely the largest, objectively measurable cost group and necessary to know for significant improvements in society and comparisons between different systems.

Only when one knows the current situation, one can improve it or compare solutions.

**Research Necessity**

The Literature Review section shows, a recent figure for the comprehensive ratio is not existent for Germany and barely existent for other jurisdictions. This article not only provides a current figure for Europe's largest economy and explains why it matters but also proposes and transparently presents a novel yet straightforward way of arriving at it.

This research chose the jurisdiction of Germany as its example because it has the advantage of using common taxation principles and ranks neither particularly high, nor particularly low in tax rankings evaluating easiness to pay taxes and tax competitiveness (PwC, 2020; Bunn, 2022). Taxpayers there can expect to neither spend a lot more nor a lot less time on taxes, for example. Those reasons suggest the chosen test subject to be a well-suited example for –and this research's results to be comparable to– many jurisdictions.

Besides, the most current comprehensive ratio for its tax system refers to the year 1984 (Hoppe, 1996, p. 41). This makes Germany a suitable research object for three more reasons: The latest value for 1984: (1) indicates the necessity for more current knowledge; (2) can serve as a ballpark confirmation of this article's results; and (3) allows for a time comparison: It gives an indication regarding the change in tax administrative efficiencies or burdens between the 1980s and 2020s.

**Knowledge Contribution**

This research project encompassed gathering data on tax revenue and administrative costs, bringing it in logical order and drawing conclusions from it. It visualises and justifies the approaches the researchers chose. It addresses possible objections readers of this research might have with counterarguments, explanations and reasoning for its novel approach.

Although all attempts to quantify total tax administrative costs can be only an estimate, this article updates public knowledge with a current value. The figure itself as well as the paths to it transparently allow for researchers to use, scrutinise and improve them. Besides, this paper discusses reasons for and implications of the results as well as possible improvements.

**Literature Review**

While plentiful data regarding taxation is available, few compilations and analyses of this data exist to determine the ratio proposed above. The last, similar calculations found for the jurisdiction of Germany are both for 1984 and both arrived at around 16%, yet one of the two values refers only to the taxpayer's side of costs (Hoppe, 1996 p. 41).

More calculations exist only for other jurisdictions and those also seldom refer to recent years. For Canada for instance, Vaillancourt (2013 p. 91) calculated 4.1-5.1% for 2007 and 4.7-5.7% for 2011 (p. 99) – when using OECD's (2023) number for Canada's tax revenue for the 2011-value.

Other available gauges are only remotely similar, as they focus on specific taxes and often either compliance cost or the government side of costs, but not for comprehensive costs of the overall tax system. Vaillancourt estimated "the total cost of operating the personal income tax and payroll tax system in Canada in 1986 [...] to be [...] 6.9 per cent of the taxes collected." (Bird, 1990). Blaufus (2019, p. 953) calculated the taxpayer side of the costs for income tax to be only 2.03-2.92% of tax revenue for Germany in 2015, despite claiming "international comparisons illustrate that the German burden is still located in the upper middle."

Again for Canada, Vishnuhadevi's (2021 Appendix) literature review mentioned the compliance cost alone for VAT to be 40% of tax revenue – referring to a study conducted by the Canadian Federation of Independent Business with 25,362 respondents for 1991. The compliance cost of personal income tax in the 1980s was mentioned in Pope (1993, p. 76) to be 2.5% for Canada, 3.6% for the UK, 5-7% for the US and 7.9-10.8% for Australia. Compliance cost of public companies' income tax was 11.4-23.7% for Australia, 2.2% for the UK and up to 19% for New Zealand (p. 77).

The variety of results arguably also indicates intrinsic uncertainty connected with calculating such values and the different methods used –as likewise suggested by York (2018)– as opposed to actual differences. Such uncertainties might be higher when trying to separate 'compliance' (meaning taxpayer cost) and government cost, all for different kinds of tax – as opposed to looking at aggregate costs and revenues.

Evans' (2007, p. 457) literary analysis "suggest[s] that compliance costs of [income and value added] taxes are typically anywhere between two percent and ten percent". The author's review further indicates that compliance costs exceed "administrative costs" –the latter meaning the cost of government– and that the average EU company with less than 250 employees has compliance costs equivalent to 31% of taxes paid in the early 2000s (p. 458). Evans further concludes that compliance costs have risen between the early 1990s and early 2000s in the US and EU (p. 459). For the US federal income tax in the year 2004, Moody (2005, p. 2) calculated compliance costs of 24.4% of tax revenue; on the rise since 1990, where it only amounted to 14.1%. On the other hand, Keen and Slemrod (2017, p. 9) –referring to a 2004 Slemrod paper– put those 'complying' costs at 11% and only 0.6% for the government side of costs.

Most available literature focuses on the government side of cost and ignores the taxpayer's ('compliance') costs, because the former is readily available and the latter are hard to approach. However, this likely distorts reality, because most jurisdictions have a 'declare yourself' system, forcing the bulk of tax administrative burden onto the taxpayer; see also Evans (2007, p. 458). The OECD (2022a, Table D.3), for example, even only focuses on "Recurrent cost of collection" –that is only regularly reoccurring costs and only the government side– where it calculates for Germany 1.2% for 2018 and 2019 and 1.4% for 2020. Therefore –albeit being plenty– these assessments seem least helpful and less relevant for this research issue, other than illustrating that the bulk of costs arise on the taxpayer side.

Methods used include most commonly quantitative questioning, where participants self-declare how much time, money or both they spend on tax administration (Vaillancourt, 2013, p. 102; Bird, 1990, p. 356; Blaufus, 2019, p. 925; Vishnuhadevi, 2021, Appendix; Pope, 1993, p. 77; Evans, 2007, p. 456). Beside questionnaire surveys, interviews, diary and case-studies, previous literature applied documentary analyses, estimating/simulating techniques and combined approaches (Evans, 2007, p. 456). Moody (2005, pp. 14–15) for example combined average hours needed to fill out forms according to the IRS with data provided by survey participants. This data was average hourly earnings of taxpayers and their tax professionals and the ratio with which tax professionals fill out said forms.

This article presents a new method, which limits its comparability with previous research but equips the research community with a novel approach for estimating tax administrative costs, in particular the tricky taxpayer side of those costs. While official government data was compiled to evaluate tax revenues and the government side of tax administrative costs, it uses tax advisers' average per person revenue, the total amount of tax advisers and the percentage of taxpayers who use their services to approach the taxpayer side of tax administrative cost.

## Methods and Materials

### Classification

As mentioned in the Introduction, this article defines tax administration cost as costs –whether measured in currency, labour or other forms– **for the revenue side of taxes**; for example for preparing documents, declaring or auditing. For further details, see Background section.

These administrative costs were divided into three categories:

1. The government's cost; especially for tax inspectors and other tax officials (category 1, Cat1).
2. The cost the taxpayer has for complying with the tax system; this can be subdivided into:

- 2.1 External costs. This is the cost taxpayers spend for outsourcing their tax duties to other firms; e.g., for tax lawyers, consultants or auditing companies (Cat2).
- 2.2 Internal costs. Those are the costs the taxpayers have for complying themselves. Because they personally fulfil their tax duties and fill out their tax forms themselves, or because their internal employees do (Cat3).

This publication gathers data on Cat1 and Cat2. It gauges Cat3 with information on the percentage of taxpayers who outsource their tax duties. It assumes: Taxpayers who use tax consultants outsource their tax administrative burden to those consultants and that their administrative costs are therefore reflected in the tax advisers' revenues. Whereas taxpayers who fill out their own tax forms –hence who do not use tax consultants– could have outsourced for the same cost. The only difference is that one is visibly measured in currency and the other is not. Arguments and counterarguments for that approach are discussed under Rationalisation.

Data had to be compiled from different sources and various assumptions were necessary to arrive at the final value. Data-gathering and -choosing; thought- and decision-making processes are explained, justified and made transparent, to allow for thorough scrutinisation.

**Tax Revenue**

The denominator is the total tax revenue for tax administered by federal and state governments: According to Destatis (2022), those were in 2021 in million euro: 833,189 - 1,098 = 832,091. Destatis is the Federal Bureau of Statistics ("Statistisches Bundesamt"), which ought to deliver accurate numbers.

Though Real estate tax ("Grundsteuer") and Commerce tax ("Gewerbesteuer") belong by law to the municipalities –GG (2022) Art 106 (6)– they are almost entirely administered by the state governments, GG (2022) Art 108 (2), (4). Hence both are included and only other municipal taxes ("Sonstige Steuern") were excluded. Destatis' (2022) tax revenue figures include all German tax revenue. Only taxes administered by federal and state governments are included, which are 99.9% (832,091/833,189) of total taxes. The 99.9%-figure includes tax that belongs to the EU – especially a fraction of value added tax. Because it too is administered by national, federal or state governments (Destatis, 2017, p. 14; 2019, 2021, 2022).

This makes sense because it is irrelevant how taxes are distributed after they were raised when determining the ratio of 'costs of raising taxes' to 'tax revenue.' The EU does not raise nor administer any taxes itself, but only receives funds. Therefore, the EU neither affects the numerator (tax administration cost) nor the denominator (raised tax).

Much like the German states administer and collect tax revenue for the German federal government, they do that too for the European Union. This article's interest lies in the ratio of total tax administrative costs (the EU does not administer tax) to the government's overall tax revenue (which includes the EU's part), and not in the ratio of any particular level of government. What is needed to fund the EU is an 'expense', not 'reduced revenue' and hence does not affect the revenue side of taxes / raised tax revenue.

**Cat1 – The Government's Costs**

Similar to tax revenue, costs for Cat1 can be determined with high confidence, because those are government expenses which are publicly documented, often even down to the cent values. The following excerpt of Table 1 illustrates how Cat1 was determined. For Table 1:

- When possible, the real (budget accounts) or at least updated values (budget plan for later years, with values for 2021 as reference) were used. Otherwise, the planned values were used (budget plan for 2021).
- The page numbers refer to the PDF-page-number of the latest reference mentioned before, which is found in the "Source" column. For example: The figure for the Federal Ministry of Finance is found on PDF-page 612 of the PDF-document Bund 2022.
- Last two columns in €; except original "Quoted 2021 value" sometimes in thousand €, especially when there is only one decimal place behind the comma.

**Table 1 (excerpt)**

*Government's tax administrative cost by position*

| EN | Original name | Source | Quoted 2021 value | In EN numerical |
|---|---|---|---|---|
| **Federal Government** | Bund | Bund 2022 | | |
| Federal Ministry of Finance | 08 Bundesministerium der Finanzen | p. 612 | 8.424.464.279,58 | 8,424,464,280 |
| of which not for tax administration: | | | | |
| Amends of the federal government | 0801 Wiedergutmachungen des Bundes | p. 615 | 1.381.688.863,89 | 1,381,688,864 |
| Federal Information Technology Center | 0816 Informationstechnikzentrum Bund | p. 674 | 944.929.863,45 | 944,929,863 |
| Financing of the successor institutions of the Treuhandanstalt | 0803 Finanzierung der Nachfolgeeinrichtungen der Treuhandanstalt | p. 626 | 375.000.209,95 | 375,000,210 |
| Burdens related to the stay or withdrawal of foreign armed forces | 0802 Lasten im Zusammenhang mit dem Aufenthalt bzw. Abzug von ausländischen Streitkräften | p. 621 | 75.153.464,27 | 75,153,464 |
| Only for tax administration: | | | | <u>5,647,691,878</u> |
| % of Federal Ministry of Finance | | | | 67.04% |
| Population in 2021, sum of 16 states | | Davies 2022 | | 83,236,000 |
| €/citizen in 2021 | | | | 67.85 |

The full Table 1 –due to its size found under Appendix– contains the above as well as the values for Germany's 16 states and finally the sum of Cat1-costs of "Federal and states' governments combined," which totals <u>17,667,088,104€</u>.

**Cat2 and 3 – The Taxpayer's Costs**

For those categories, it is important to note the following: Being a 'declare yourself' system, the German taxation system passes the bulk of tax administrative burden onto the taxpayer. Taxpayers need to find out themselves what taxes they need to pay, what forms to fill out, how to correctly declare and determine their tax burdens and prepare and present all the information leading to those determinations to the government. The state then only audits with software. And a tiny fraction of tax declarations is additionally audited with human labour.

Generally, tax officials ought to only help taxpayers free of charge in cases of a "clear […] error" or when answering questions regarding administrative, procedural processes (DE-AO, 2017 Section 89 (1)). Therefore, the state is only involved in a small part of the total tax administrative burden. Because the overwhelming majority of taxpayers fulfil their tax duties themselves (Mülhens, 2016) –as opposed to hiring a contractor to fulfil those duties for them, such as a tax consultant (outsourcing)– the larger portion of taxpayer's tax administrative cost cannot be measured in money spent and must be gauged.

In this task, the following is helpful: In the jurisdiction of Germany, generally only certain people are allowed to conduct "commercial assistance in tax matters", which are "tax consultants, tax agents, lawyers, established European lawyers, auditors and chartered accountants" (DE-StBerG, 2022, §3). Especially tax consultants must therefore be chartered members of the Chamber of Tax Advisors to offer tax consulting services for money, but also to carry the title 'Tax consultant' ("Steuerberater"). Members of the chamber must pay a regular fee of around 500€ p.a., depending on the chamber, and maintain special business insurance, which starts at over 100€ p.a. and increases with the size of the operation, to be listed as chamber members. In addition, they have to deal with the tax chamber administratively to maintain membership and likely have as formal members increased tax duties themselves, too. Therefore, it is presumed that most if not practically all members are active, because otherwise no meaningful advantages justify the monetary and administrative disadvantages that come with being a formal member.

Despite the name, tax consultants seldom purely advise or 'consult', but often rather fulfil many of their client's tax duties. Since one knows the number of active tax consultants (BStBK, 2021) and the amount of revenue per tax consultant (Juve, 2022; BStBK, 2019), one can easily calculate Cat2.

Yet, the most challenging category, with admittedly the highest uncertainty, is Cat3. It is known that 28% of German taxpayers use a tax consultant (Mülhens, 2016), hence outsource their tax administrative burden. That means the other 72% do their tax duties themselves, resulting in internal (opportunity) cost that is not formally accounted for.

So this publication's approach presumes those hidden, internal costs to be equal to the costs that would be visible, if they were outsourced, resulting in Cat3 to be $(1/0.28) \cdot \text{Cat2} - \text{Cat2}$.

*Rationalisation*

Albeit this being the most straightforward and hence the most unbiased, impartial and objective approach available, some readers may be sceptical about it. The following list of possible reasons why the chosen approach might be off from reality attends this possible, first-glance scepticism. Within every argument (marked with an "a" after its number) is the symbol [+], if the specific point indicates the approach to be an overcount of costs. Or the symbol [-], in case the point indicates an undercount. A counterargument (b) may follow an argument.

**1a:** Tax consultants are not the only recipients of outsourced tax administrative costs [-]. As shown above, DE-StBerG (2022) §3 gives lawyers, accountants, auditors and more also permission to charge for tax services, hence tax administrative cost is outsourced to them also. Further, there is a whole list of other persons in DE-StBerG (2022) §4 –including 'wage tax assistance unions' („Lohnsteuerhilfevereine")– all creating significant, visible revenue for outsourced tax administrative costs not present in Cat2. By Cat2 being an undercount, Cat3 will also be proportionally. So one would need to add their revenue to the revenue of tax consultants to have the full, outsourced administrative cost for taxation (=Cat2), resulting also in a more accurate Cat3.

**1b, counterpoint:** That is true. Yet their cost is harder to measure, especially which of their revenue is precisely for tax services. Besides, the people using their services are mostly not present in the 28%, so adding other tax services' revenue to that of tax consultants would lead to an overcount. Tax consultants have the advantage of dealing overwhelmingly with tax, making their revenue less 'polluted' with non-tax-related revenue, which mostly is not true for the other entities and groups of persons mentioned above in 1a. But tax consultants also do service necessary for non-tax-issues, like transparency –e.g. accounting and publishing of company accounts– albeit to a minor degree. The undercount mentioned in 1a is at least partly offset by the current overcount due to that.

**2a:** There are plenty industries' subsectors currently not accounted for, which create revenue from tax administration related services [-]. Those include literature about tax (of which there is an enormous amount, especially for Germany's tax system), schools and research institutions (for-profit and government-run), software for Germany's tax system (of ever-increasing value) and many more.

**2b, counterpoint:** This is also true. This irrefutably indicates an undercount of the current approach. Yet it also is not an undercount by 100%, because those are often input costs for the revenue of tax consultants. As far as that is the case, revenue of the subsectors mentioned in 2a (which *is* tax administrative cost for the economy) is input cost for (and therefore already incorporated and reflected in) tax consultants' revenue. Again, trying to distinguish and finding information about the parts of revenue of those subsectors, which are related to tax services, is impractical to impossible – and what is even harder: Determining which of that revenue is not yet accounted for with tax consultants' revenue, by being tax consultants' input costs. Therefore, trying to be more precise here with trifles will likely result in disimprovement.

**3a:** Taxpayers with more tax duties –hence higher tax administrative costs– tend to be more likely to use tax consultants as opposed to taxpayers with fewer tax administrative burdens. Assuming them to be proportional will therefore lead to an overcount [+].

**3b, counterpoint:** While one might instinctively imagine that, it does not seem to be true; at least when one presumes higher income to correlate with higher administrative tax burdens. In Mülhens' 2016 survey of over 1,000 participants, households with monthly net income of ≥3,000€ used a tax consultant only in 22.3% of cases –distinctly below average– while both groups with lower income (2,000-3,000€ and 2,000-2,500€) scored more than 10 percentage points higher.

Besides, at least when looking at the corporate sector, the opposite seems logical: When a corporation is large enough, it makes more sense for it to have internal tax departments, which do a lot, if not all the tax work. Those employees are not required to be members of the Chamber of Tax Advisors, which can result in a significant undercount in the current approach. It is essentially a 'make-or-buy-decision'. With size shrinks the relative overhead. The higher tax burdens, the more it then makes sense to 'make', hence to handle tax duties in-house.

Until there is objective information indicating disproportionality, it is the more impartial, objective and the default decision to expect proportionality. In the case at hand, expecting proportionality is even more appropriate, because there is logic and empirical information opposing the disproportionality thesis.

Though, admittedly, disproportionality cannot be ruled out. The logical evidence to refute it may seem weak to some and the empirical evidence is only indicative and from a single source. Also, the whole estimation of Cat3 hinges on the same, single, likely non-peer-reviewed, commercial market research study, assisted by the mentioned logic. That's why the authors emphasise the high degree of uncertainty when trying to quantify Cat3.

**4a:** Even those who have a tax consultant are unlikely –or it is even impossible– to be entirely free of tax administrative work. Often, one needs to prepare documents for tax consultants, attend meetings with them, with tax authorities etc; resulting in an undercount in the approach [-].

**4b, counterpoint:** This is a valid point reflective of reality, leading undoubtedly to an undercount in the current approach. The only counterargument is that it is also here hard –if not impossible– to accurately account for those hidden (opportunity) costs, which is why the authors avoided trying that. Though, future researchers may try to measure those costs, for example with surveys or field studies.

Considering all those points, the chosen, following calculation of costs is more likely to be an undercount than an overcount. Though, readers are reminded that Cat3 is merely an estimation with a high degree of uncertainty. The approximation is based on assumptions, which are only backed up by indicative evidence and heavily relies on Mülhens' (2016) results. Above text –between Cat2 and 3 and Rationalisation– and the following Table 2 transparently describe how Cat2 and Cat3 were derived from the compiled data.

**Table 2**
*Calculations for Cat2 and 3*

| Description | Source | [1] | Notes | Value |
|---|---|---|---|---|
| Revenue per CTAm[2] in € | Juve 2022 | m | [3] | 482,339 |
| Alternative value: | | | | |
| Revenue per CTAm with sole office (2017) in € | BStBK 2019 | g | [4] | 332,000 |
| Inflation multiplier from 2017 to 2021 | Bundesbank 2022 | g | [5] | 1.082846 |
| 2017-value adjusted for inflation in € | | | | 359,505 |
| Amount of CTAm on 2021-01-01 | BStBK 2021 | g | [6] | 100,204 |
| -> SubtotalCat2; 482,339€ * 100,204 = € | | | | 48,332,295,152 |
| Amount of taxpayers who outsource their tax duties | Mülhens 2016 | m | [7] | 0.28 |
| -> TotalCat2+3 in €: | | | [8] | 172,615,339,828 |
| Lower bound in €: | | | [9] | 128,656,522,584 |
| Ø of above two values in €: | | | | 150,635,931,206 |
| Final value for Cat2&Cat3 in €: | | | | 150,635,931,206 |

[1] = Expected quality/reliability of source or value (g=good, m=moderate, p=poor).
[2] = Chamber of Tax Advisors member(s).
[3] = Market data analysis: Average revenue per professional title holder of the top 30 tax consulting firms in Germany for 2020 and 2021, whereas for 2020 there is one value missing. 482,339€ is the sum of the 59 values divided by 59. Though: Full Time Equivalents. So the real number is likely lower. This was addressed by lowering the final value, through taking the mean of that figure and the alternative value.
[4] = Alternative value for above. Reliable official, source, though values from 2017 and for sole trading professional title holders only. So overall, not as appropriate as above value.
[5] = Reliable, official source. Though, CPI-inflation-adjustment is a simplified approach to adapt the 2017-values to 2021. 2021-12-value/2017-12-value = 111.1/102.6.
[6] = Reliable, official source.
[7] = Moderately reliable source; a survey of 1002 participants from 2016.
[8] = SubtotalCat2 divided by 0.28, because only 28% of taxpayers outsource their tax duties, whereas the others need to pay with their own time or with currency by hiring internal employees. Arriving at Cat2+Cat3.
[9] = Using 2017-value adjusted for inflation instead of 2020/2021-Ø-value of top 30 tax consulting firms.

# Results

## Results for Germany in 2021

Total tax revenue, net of tax administered by municipalities (Ttrnotabm):  832,091,000,000€

| | |
|---|---|
| **Cat1:** | **17,667,088,104€** |
| Cat2 (primary value): | 48,332,295,152€ |
| Cat2 (alternative value): | 36,023,826,324€ |
| Cat3 (primary value): | 124,283,044,676€ |
| Cat3 (alternative value): | 92,632,696,261€ |
| Cat2+3 (primary value): | 172,615,339,828€ |
| Cat2+3 (alternative value): | 128,656,522,584€ |
| **Cat2+3 (mean):** | **150,635,931,206€** |
| | |
| Cat1+Cat2+3 (mean): | 168,303,019,310€ |
| Cat1+Cat2+3 (mean)/Ttrnotabm: | 20.23% |
| When including tax administered by municipalities, the ratio would be: | 20.20% |

## Summary of Results - Comparison with Expectations and Existing Literature

This research utilised a new method for calculating tax administrative costs. Hence, above results have limited comparability with pre-existing research. Despite this, the final values align with what other researchers found using their methods.

Only around 10% of the total administrative cost arises on the government side, while around 90% is taxpayer's cost. This is in line with expectations, as Germany –like most jurisdictions– demands from the taxpayers the bulk of administrative burden, such as to calculate, prepare and declare their taxes themselves.

Also, the roughly 2% –government administrative cost as a percent of tax revenue– aligns with the OECD's (2022a, Table D.3) estimations of 1.4% for 2020, because the OECD's number only considers "recurrent" cost of government; as also mentioned under Literature review.

The overall results match with Moody's (2005, p. 2) report, who calculated the compliance cost for US federal income tax to be 24.4% of tax revenue for the year 2004, drastically up from 14.1% in the year 1990.

Compliance cost at ca 18% of tax revenue is between Blaufus' (2019, p. 953) gauge of around 2.5% for income tax in Germany for 2015, Keen's (2017 p. 9) mentioning of 11% for the US for the early 2000s – and Vishnuhadevi's (2021 Appendix) 40% for VAT in Canada for 1991 or Evans' (2007, p. 458) 31% of total taxes paid for EU companies with <250 employees for the early 2000s.

A comparison with Hope (1996, p. 41) indicates a historical rise in tax administrative cost as percentage of tax revenue for Germany between 1984 and 2021. Moody (2005, p. 2) and Evans (2007, p. 459) too found compliance cost to be on the rise, at least between the early 1990s and early 2000s in US and EU; while Blaufus (2019, p. 928) claimed income tax costs have decreased between 2007 and 2015 for Germany – at least for "employees who self-prepare their tax returns [...], even though the number of tax returns increased."

**Discussion**

This article shows that the ratio of tax administrative costs to tax revenue is likely to be over 20%, for Germany in the year 2021. This indicates: higher tax administrative burdens outpaced possible productivity gains from digitalisation. It also shows the unsustainability of the path Germany's tax framework is currently on.

A high rise in tax burdens accompanied with starting a business hinders entrepreneurship and a startup culture, hampering progress and innovation. Due to lower amortisation of fixed costs through scaling, tax administrative burdens are likely to hit startups and SME's especially hard; evidence for this is also found by Evans (2007, p. 458).

While one can be very confident in values for Cat1 and somewhat confident in values for Cat2, uncertainties arise especially for Cat3. Other researchers are invited to propose better solutions to measure in particular that category of administrative tax burdens. Also, research for other tax systems would allow for comparisons between different jurisdictions, but also with other data, for example with PwC's (2020) research regarding "Time to comply."

**Possible Explanations**

Reasons for the current, arguably high administrative costs and the indication of a worsening of administrative burden, despite major advancements in digitalisation and automation capabilities between 1984 and 2021, might include the following:

*Historical Reasons*

The first tax systems were based on property, especially ownership of agricultural land and animals, upon which taxes were levied proportionally. This was because most people owned small plots of land which they cultivated.

During and after industrialisation, the majority of people moved from self-owned, small farms to the cities into factories, no longer owned land but had salary income. Tax regimes changed to focus on income. Current tax systems are mostly still built on the factory worker model, despite the largest wealth gaps today being no longer in salary or income, but in ownership. Monitoring cash flow, revenue, expenses and income is vastly more burdensome than taking an annual snapshot of ownership.

Instead of fundamental change, 'change' constituted mostly additions and higher complexity on top of the existing system, which seems to have made it fussier and more cumbersome between 1984 and 2021. Though ownership gaps outpaced income gaps (Bartels, 2022, Figure 3; World Bank, 2023), a focus on ownership instead of income did not occur, which could have increased tax efficiency and redressed administrative strain.

*Political Reality*

Structural and fundamental changes in taxation –which would be necessary to leave the current path and sustainably reduce tax administrative burden– face a high political hurdle yet to be overcome by most jurisdictions: They seem risky. Politicians may be too risk-averse and do not want the responsibility to oversee such drastic, uncharted change.

Politicians may also lack the multidisciplinary competence and confidence to fully understand and reform taxation, which includes economics, law, philosophy –including logics and ethics– psychology, sociology, mathematical statistics etc.

Hence most jurisdictions' legislators rely on outside professionals, who draft tax code changes for them. Those often include large accounting and consulting firms, who have an invested interest in maintaining the current system and for whom more tax administrative burden means more business. Legislators might hardly find law or tax professionals without such conflicts of interest. If laws and tax were to become simple, professionals and their acquired knowledge, position and advantage would become obsolete. Those might further explain why worsening instead of improvement occurred between 1984 and 2021.

*Naive Interventions* (a term made popular by Taleb [2012, Chapter 7], describing disimprovement)

Depicting life's full chaotic complexity and account for it in a tax code requires a lot of legal text and administrative effort. Politicians may believe increased tax code complexity is necessary for more justice, because it more accurately accounts for the eventualities and complexities in life. Tax also often tries to steer society away from unwanted and towards perceived desired activity.

Yet, the outcome is likely not only to result in more administrative burden, but also less –not more– tax justice: Those with resources can use the tax code's complexity, while those with less resources –money, time, access to and connection with professionals etc– cannot. Over time, this tends to result in longer tax codes, higher administrative burdens, and less social justice. Apart from that, decisions might no longer be made based on real-world logic, but on tax considerations; especially when there are steering attempts. So, tax code complexity will not only lead to more administrative burdens, social injustice and wealth gaps, but also to economically imperfect allocation of resources and activity.

*Many Tax Forms as Opposed to One*

Different kinds of taxes have been introduced over time, increasing the number of different forms of tax. This may be, because this makes the total tax burden less transparent to the taxpayer. One single, consolidated tax would be administratively easier, but also more transparent; which is of course socially, ethically and economically positive, yet may make it politically less acceptable. The OECD (2022b, Table 1.1) estimates Germany's tax burden to be 39.5% of GDP for 2021; Eurostat (2022) ascertains 42.4%. The latter includes net social contributions and is close to the EU and Euro area averages of 41.7% and 42.2% (Eurostat, 2022). If the government demanded openly and transparently 40% of the average person's added value, the taxpayer might not accept that. This might contribute to why Germany stayed with an untransparent, flawed system –despite better alternatives available– and the lack of improvement between 1984 and 2021.

**Possible Solutions**

A fundamental change in the taxation framework, tailored to the modern times, may be necessary. This includes addressing the social reality –where ownership as opposed to income may be more focused– as well as designing it around modern automation capabilities. One form of tax replacing the many currently in place may also reduce tax administrative burden, as for instance also proposed and shown by Tuerck (2007).

Estonia is an example proving that it is possible to overcome the impeding reasons mentioned above and implement some of the outlined possible solutions. It shows transparent, digitalised and simplified taxation is not only possible, but also successful in achieving low administrative burden –such as measured in hours needed to comply with tax duties (PwC, 2020)– while also outperforming in other metrics typically connected with efficient and beneficial taxation and prudent fiscal policy: such as tax competitiveness, where it ranks first within OECD countries, or government debt to GDP, which stands in 2023 at 19.4% for Estonia; by far the lowest in the EU and Euro area, where the averages are 85% and 91.3% (Bunn, 2022; IMF, 2023).

Another idea might be to outsource –even more– government costs to the taxpayers. Though, in the prevalent self-assessment system that most jurisdictions use, the majority of tax administrative cost is already located on the taxpayer side. In the case of Germany, that is roughly 90%, see under Results. Outsourcing the government's duties further –for example for auditing or collection– to profit-oriented businesses –for example via a bidding process– is imaginable, yet might constitute a conflict of interest and endanger the rule of law. One reason for that is, companies would be interested in cost reduction while the government is interested in thorough and fair auditing. Besides, since only around 10% of the cost arises on the government's side, it might be wiser to first focus most directly on reducing the 90% instead.

**Conclusions**

This article illustrates, Germany's ratio of the revenue side of tax administrative cost to total tax revenue likely exceeded 20% in the year 2021. Though there is significant uncertainty, in particular regarding the quantification of hidden (opportunity) costs for taxpayers fulfilling their own tax duties, the Rationalisation-chapter lays out why the calculations for this category –and hence for overall costs– are more likely to be an undercount than an overcount.

The findings indicate a noticeable rise in this ratio since 1984, where researchers estimated it to be around 16% (Hoppe, 1996, p. 41) – despite computers, internet, digitalisation, AI and other automation capabilities since then. This insight is significant for public policy decision-making, as it indicates need for fundamental change in Germany's taxation system and –due to its comparability– for other jurisdictions too. Though, the new method presented in this article limits the comparability of its results with previous research. Possible reasons for high administrative costs, likely implications of it and feasible solutions were also outlined.

Acknowledgements

The authors give special thanks to Prof. Dr. Yuri Biondi for suggestions in earlier drafts and sincerely thank peer-reviewers, editors and staff of the *Journal of Multidisciplinary Research,* ISSN: 1947-2900 (print), 1947-2919 (online), –where this article has been originally published in an earlier version in 2024-09– for their feedback and enabling this article to be released.

# Appendix

- Table 1 (full): Government's tax administrative cost by position [XLS]; yet also pasted below: arweave.net/AM9i75AtQtThIeTxKBFsHP2fVghZzB7E8tj20UtEm7U

- Sources list with more details, human and machine readable [TXT]: pastes.io/raw/d7a3gbh8jx or pastejustit.com/raw/hwvuohiij0

Above files were both archived on 2023-09-11 at:

- web.archive.org/web/20230911/arweave.net/AM9i75AtQtThIeTxKBFsHP2fVghZzB7E8tj20UtEm7U

- archive.md/qS8Rt

**Table 1 (full)**
**Government's tax administrative cost by position**

| EN | Original name | Source | Quoted 2021 value | In EN numerical |
|---|---|---|---|---|
| **Federal Government** | Bund | Bund 2022 | | |
| Federal Ministry of Finance | 08 Bundesministerium der Finanzen | p. 612 | 8.424.464.279,58 | 8,424,464,280 |
| *of which not for tax administration:* | | | | |
| Amends of the federal government | 0801 Wiedergutmachungen des Bundes | p. 615 | 1.381.688.863,89 | 1,381,688,864 |
| Federal Information Technology Center | 0816 Informationstechnikzentrum Bund | p. 674 | 944.929.863,45 | 944,929,863 |
| Financing of the successor institutions of the Treuhandanstalt | 0803 Finanzierung der Nachfolgeeinrichtungen der Treuhandanstalt | p. 626 | 375.000.209,95 | 375,000,210 |
| Burdens related to the stay or withdrawal of foreign armed forces | 0802 Lasten im Zusammenhang mit dem Aufenthalt bzw. Abzug von ausländischen Streitkräften | p. 621 | 75.153.464,27 | 75,153,464 |
| Only for tax administration: | | | | <u>5,647,691,878</u> |

| | | | |
|---|---|---|---|
| % of Federal Ministry of Finance | | | 67.04% |
| Population in 2021, sum of 16 states | | Davies 2022 | 83,236,000 |
| €/citizen in 2021 | | | 67.85 |

| | | | | |
|---|---|---|---|---|
| **North Rhine-Westphalia** | Nordrhein-Westfalen | Nordrhein-Westfalen 2022 | | |
| Ministry of Finance | Ministerium der Finanzen | | 2.803.203.700 | 2,803,203,700 |
| of which not for tax administration: | | | | |
| State building administration – Upper financial office NRW | Staatliche Bauverwaltung Oberfinanzdirektion NRW | | 8.718.700 | 8,718,700 |
| Asset management after liquidation of special assets | Vermögensverwaltung nach Auflösung von Sondervermögen | | 4.442.700 | 4,442,700 |
| General allowances | Allgemeine Bewilligungen | | -11.303.100 | -11,303,100 |
| Only for tax administration: | | | | <u>2,801,345,400</u> |
| % of Ministry of Finance | | | | 99.93% |
| Population in 2021 | | Davies 2022 | | 17,925,000 |
| €/citizen in 2021 | | | | 156.28 |
| **Bavaria** | Bayern | Bayern 2021 | | |
| State Ministry of Finance and Homeland | 06 Staatsministerium der Finanzen und für Heimat | p. 171 | 3.032.230,3 | 3,032,230,300 |
| of which not for tax administration: | | | | |
| Administration of Government Castles, Gardens and Lakes | 06 16 Verwaltung der staatl. Schlösser, Gärten und Seen | p. 122 | 168.851,8 | 168,851,800 |
| State Ministry of Security in Information Technology | 06 20 Landesamt für Sicherheit in der Informationstechnik | p. 135 | 21.387,5 | 21,387,500 |
| State Ministry of Digitalisation, Broadband and Surveying | 06 21 Landesamt für Digitalisierung, Breitband und Vermessung | p. 155 | 161.695,0 | 161,695,000 |
| Offices for Digitalisation, Broadband and Surveying | 06 22 Ämter für Digitalisierung, Breitband und Vermessung | p. 163 | 132.842,8 | 132,842,800 |
| pure non-tax admin total: | | | | 484,777,100 |
| <u>neutral:</u> | | | | |
| Bavaria-Servers and state communication infrastructure | 06 50 Bayern-Server und staatliche Kommunikationsinfrastruktur | p. 170 | 8.000,0 | 8,000,000 |
| Ministry | 06 01 Ministerium | p. 14 | 49.002,0 | 49,002,000 |
| Other costs of subcategory 06 / Accumulations for the category of Epl. 06 | 06 02 Sammelansätze für den Gesamtbereich des Epl. 06 | p. 26 | 764.228,5 | 764,228,500 |
| neutral total | | | | 821,230,500 |
| | | | | |
| Only pure tax administrative costs: | | | | 1,726,222,700 |
| Ratio pure-tax to (pure-tax & pure-non-tax): | | | | 78.07% |
| | | | | |
| 'Pure tax admin cost' + 'Ratio pure-tax to (pure-tax & pure-non-tax)' * 'neutral costs': | | | | <u>2,367,392,695</u> |
| % of State Ministry of Finance and Homeland | | | | 78.07% |
| Population in 2021 | | Davies 2022 | | 13,177,000 |
| €/citizen in 2021 | | | | 179.66 |
| **Baden-Wuerttemberg** | Baden-Württemberg | Baden-Württemberg 2022 | | |
| Ministry of Finance | 06 Ministerium für Finanzen | p. 3 | 1.779.803,1 | 1,779,803,100 |
| <u>of which not for tax administration:</u> | | | | |
| State Statistical Office | 0607 Statistisches Landesamt | p. 52 | 59.364,1 | 59,364,100 |
| State Center for Data Processing | 0610 Landeszentrum für Datenverarbeitung | p. 74 | 101.327,9 | 101,327,900 |
| Federal Building Baden-Württemberg | 0614 Bundesbau Baden-Württemberg | p. 79 | 0,0 | 0 |
| Assets and construction Baden-Württemberg | 0615 Vermögen und Bau Baden-Württemberg | p. 89 | 157.633,3 | 157,633,300 |
| Companies and holdings | 0620 Betriebe und Beteiligunge | p. 117 | 32.344,8 | 32,344,800 |
| State coins of Baden-Württemberg | 0622 Staatliche Münzen Baden-Württember | p. 126 | 0,0 | 0 |
| Wilhelma in Stuttgart-Bad Cannstatt | 0623 Wilhelma in Stuttgart-Bad Cannstatt | p. 133 | 11.342,6 | 11,342,600 |
| Meersburg State Winery | 0624 Staatsweingut Meersburg | p. 141 | 0,0 | 0 |
| Only for tax administration: | | | | 1,417,790,400 |
| % of Ministry of Finance | | | | 79.66% |
| Population in 2021 | | Davies 2022 | | 11,125,000 |
| €/citizen in 2021 | | | | 127.44 |
| **Lower Saxony** | Niedersachsen | Niedersachsen 2022 | | |
| Ministry of Finance | 04 Finanzministerium | p. 228 | 1.055.461.836,74 | 1,055,461,837 |
| of which not for tax administration: | | | | |
| State construction management Lower Saxony – budgeted | 0410 Staatliches Baumanagement Niedersachsen - budgetiert | p. 228 | 252.454.358,58 | 252,454,359 |
| Lower Saxony State Property Fund -Fund Management- | 0440 Landesliegenschaftsfonds Niedersachsen - Fondsverwaltung- | p. 228 | 4.232.982,06 | 4,232,982 |
| Only for tax administration: | | | | 798,774,496 |
| % of Ministry of Finance | | | | 75.68% |
| Population in 2021 | | Davies 2022 | | 8,027,000 |
| €/citizen in 2021 | | | | 99.51 |

| Hesse | Hessen | Hessen 2021 | | |
|---|---|---|---|---|
| Hessian Ministry of Finance | 06 Hessisches Ministerium der Finanzen | p. 205 | 1.123.556.100 | 1,123,556,100 |
| of which not for tax administration: | | | | |
| Hessian lottery administration | 06 12 Hessische Lotterieverwaltung | p. 205 | — | 0 |
| State Office for Construction and Real Estate Hesse | 06 13 Landesbetrieb Bau und Immobilien Hessen | p. 205 | 5.907.000 | 5,907,000 |
| Hessian Competence Center for new management control | 06 16 Hessisches Competence Center für Neue Verwaltungssteuerung | p. 205 | 83.653.700 | 83,653,700 |
| Only for tax administration: | | | | 1,033,995,400 |
| % of Ministry of Finance | | | | 92.03% |
| Population in 2021 | | Davies 2022 | | 6,295,000 |
| €/citizen in 2021 | | | | 164.26 |

| Rhineland-Palatinate | Rheinland-Pfalz | Rheinland-Pfalz 2023 | | |
|---|---|---|---|---|
| Ministry of Finance | 04 Ministerium der Finanzen | p. 5 | 609.554.045 | 609,554,045 |
| of which not for tax administration: | | | | |
| Federal Building Office | 04 08 Amt für Bundesbau | p. 105 | 5.285.469 | 5,285,469 |
| State Construction Administration | 04 10 Staatliche Bauverwaltung | p. 114 | 0 | 0 |
| Benefits under the Federal Compensation Act (BEG) | 04 14 Leistungen nach dem Bundesentschädigungsgesetz (BEG) | p. 119 | 33.559.555 | 33,559,555 |
| Compensation Administration | 04 15 Wiedergutmachungsverwaltung | p. 127 | 1.284.389 | 1,284,389 |
| Structure and Approval Directorate North (SGD Nord) | 04 80 Struktur- und Genehmigungsdirektion Nord (SGD Nord) | p. 142 | 1.019.360 | 1,019,360 |
| Structure and Approval Directorate South (SGD South) | 04 81 Struktur- und Genehmigungsdirektion Süd (SGD Süd) | p. 148 | 836.888 | 836,888 |

| | | | |
|---|---|---|---|
| Only for tax administration: | | | <u>567,568,384</u> |
| % of Ministry of Finance | | | 93.11% |
| Population in 2021 | | Davies 2022 | 4,106,000 |
| €/citizen in 2021 | | | 138.23 |

| | | | | |
|---|---|---|---|---|
| **Saxony** | Sachsen | Sachsen 2022 | | |
| State Ministry of Finance | 04 Staatsministerium der Finanzen | p. 35 | 557.080.770,23 | 557,080,770 |
| of which not for tax administration: | | | | |
| State enterprise Saxon real estate and construction management | 04 11 Staatsbetrieb Sächsisches Immobilien- und Baumanagement | p. 30 | 79.585.820,67 | 79,585,821 |

| | | | |
|---|---|---|---|
| Only for tax administration: | | | <u>477,494,950</u> |
| % of Ministry of Finance | | | 85.71% |
| Population in 2021 | | Davies 2022 | 4,043,000 |
| €/citizen in 2021 | | | 118.10 |

| | | | | |
|---|---|---|---|---|
| **Berlin** | Berlin | Berlin 2022 | | |
| Financial/Tax administration | 06 Finanzverwaltung | p. 11 | 470.774.100 | 470,774,100 |
| of which not for tax administration: | | | | |
| none found. | | | | |

| | | | |
|---|---|---|---|
| Only for tax administration: | | | <u>470,774,100</u> |
| % of Ministry of Finance | | | 100.00% |
| Population in 2021 | | Davies 2022 | 3,677,000 |
| €/citizen in 2021 | | | 128.03 |

| **Schleswig Holstein** | Schleswig-Holstein | Schleswig-Holstein 2022 | | | |
|---|---|---|---|---|---|
| Financial/Tax administration | 06 Finanzverwaltung | | p. 58 | 249.965,7 | 249,965,700 |

of which not for tax administration:
none found.

| | | | |
|---|---|---|---|
| Only for tax administration: | | | 249,965,700 |
| % of Ministry of Finance | | | 1.00 |
| Population in 2021 | | Davies 2022 | 2,922,000 |
| €/citizen in 2021 | | | 85.55 |

| **Brandenburg** | Brandenburg | Brandenburg 2022 | | | |
|---|---|---|---|---|---|
| Ministry of Finance and for Europe | 12 Ministerium der Finanzen und für Europa | | p. 48 | 411.805.159,00 | 411,805,159 |

of which not for tax administration:

| | | | | | |
|---|---|---|---|---|---|
| European Affairs and International Relations | 12 060 Europaangelegenheiten und internationale Beziehunge | | p. 692 | 3.457.681,77 | 3,457,682 |
| Interreg programmes | 12 065 Interreg-Programme | | p. 706 | 24.769.635,64 | 24,769,636 |

| | | | |
|---|---|---|---|
| Only for tax administration: | | | 383,577,842 |
| % of Ministry of Finance | | | 93.15% |
| Population in 2021 | | Davies 2022 | 2,538,000 |
| €/citizen in 2021 | | | 151.13 |

| **Saxony-Anhalt** | Sachsen-Anhalt | Sachsen-Anhalt 2022 | | | |
|---|---|---|---|---|---|
| Ministry of Finance | 04 Ministerium der Finanzen | | p. 298 | 281.208.953,04 | 281,208,953 |

of which not for tax administration:
none found.                                   p. 300

| | | | | |
|---|---|---|---|---|
| Only for tax administration: | | | | 281,208,953 |
| % of Ministry of Finance | | | | 100.00% |
| Population in 2021 | | Davies 2022 | | 2,169,000 |
| €/citizen in 2021 | | | | 129.65 |
| **Thuringia** | Thüringen | Thüringen 2022, Band 1 | | |
| Thuringian Ministry of Finance | 06 Thüringer Finanzministerium | p. 9 | 193.953.130,24 | 193,953,130 |
| of which not for tax administration: | | Thüringen 2022, Band 2a | | |
| Thuringia Central Transport Service | 06 20 Zentraler Fahrdienst Thüringen | p. 554 | 1.089.516,42 | 1,089,516 |
| Only for tax administration: | | | | 192,863,614 |
| % of Ministry of Finance | | | | 99.44% |
| Population in 2021 | | Davies 2022 | | 2,109,000 |
| €/citizen in 2021 | | | | 91.45 |
| **Hamburg** | Hamburg | Hamburg 2022 | | |
| Section 9.1 Financial Authority | Einzelplan 9.1 Finanzbehörde | | | 469,706,732 |
| consisting of the following components: | | | | |
| components which not for tax administration: | | | | |
| Task area 279 "Senate assistance" | Aufgabenbereich 279 „Senatsassistenz" | p. 11 | | 82,285,044 |
| 1st subcomponent | | | 17.222.787,34 | 17,222,787 |
| 2nd subcomponent | | | 47.419.811,93 | 47,419,812 |
| 3rd subcomponent | | | 7.035.515,26 | 7,035,515 |
| 4th subcomponent | | | 10.606.929,92 | 10,606,930 |
| Task area 280 "real estate management" | Aufgabenbereich 280 „Immobilienmanagement" | p. 12 | | 32,153,487 |
| 1st subcomponent | | | 22.471.864,58 | 22,471,865 |
| 2nd subcomponent | | | 2.841.338,66 | 2,841,339 |
| 3rd subcomponent | | | 6.840.284,20 | 6,840,284 |

components which only for tax administration:

| | | | | |
|---|---|---|---|---|
| Task area 278 "Control and Service" | Aufgabenbereich 278 „Steuerung und Service" | p. 10 | 18.681.359,57 | 18,681,360 |
| Task area 281 "Taxation" | Aufgabenbereich 281 „Steuerwesen" | p. 13 | | 336,586,840 |
| 1st subcomponent | | | 46.642.402,37 | 46,642,402 |
| 2nd subcomponent | | | 269.018.124,47 | 269,018,124 |
| 3rd subcomponent | | | 20.926.313,61 | 20,926,314 |

| | | | |
|---|---|---|---|
| Only for tax administration: | | | 355,268,200 |
| % of Ministry of Finance | | | 75.64% |
| Population in 2021 | | Davies 2022 | 1,854,000 |
| €/citizen in 2021 | | | 191.62 |

| | | | | |
|---|---|---|---|---|
| **Mecklenburg-Western Pomerania** | Mecklenburg-Vorpommern | Mecklenburg-Vorpommern 2022 | | |
| Department of Treasury | 05 Geschäftsbereich des Finanzministeriums | p. 7 | 245.899,8 | 245,899,800 |

of which not for tax administration:

| | | | | |
|---|---|---|---|---|
| State building and property offices | 0505 Staatliche Bau- und Liegenschaftsämter | p. 7 | 33.189,0 | 33,189,000 |
| State palaces, gardens and art collections | 0506 Staatliche Schlösser, Gärten und Kunstsammlungen | p. 7 | | 0 |
| Measures of the MV protection fund | 0580 Maßnahmen des MV-Schutzfonds | p. 7 | -- | 0 |

| | | | |
|---|---|---|---|
| Only for tax administration: | | | 212,710,800 |
| % of Ministry of Finance | | | 86.50% |
| Population in 2021 | | Davies 2022 | 1,611,000 |

€/citizen in 2021 … 132.04

| Saarland | Saarland | | Saarland 2022 | |
|---|---|---|---|---|
| **Saarland** | | | | |
| Ministry of Finance and Europe / Ministry of Finance and Science | 04 Ministerium für Finanzen und Europa / Ministerium der Finanzen und für Wissenschaft | p. 24 | 120.152.931,40 | 120,152,931 |
| of which not for tax administration: | | | | |
| State Office for Central Services - Office for Construction and Real Estate | 0412 Landesamt für Zentrale Dienste - Amt für Bau und Liegenschaften | p. 257 | 0,00 | 0 |
| State Office for Central Services - Statistical Office | 0413 Landesamt für Zentrale Dienste - Statistisches Amt | p. 259 | 7.328.100,00 | 7,328,100 |
| Promotion of science and universities | 0414 Förderung von Wissenschaft und Hochschulen | none | | |
| University of Technology and Economics | 0415 Hochschule für Technik und Wirtschaft | none | | |
| University | 0416 Universität | none | | |
| of which only for tax administration: | | | | |
| Ministry of Finance and Science | 0401 Ministerium der Finanzen und für Wissenschaft | p. 242 | 10.158.547,85 | 10,158,548 |
| General permits | 0402 Allgemeine Bewilligungen | p. 244 | 96.689,87 | 96,690 |
| Tax offices | 0404 Finanzämter | p. 250 | 89.923.982,49 | 89,923,982 |
| State Office for Central Services | 0411 Landesamt für Zentrale Dienste | p. 255 | 12.645.611,19 | 12,645,611 |

| | | | |
|---|---|---|---|
| Only for tax administration: | | | <u>112,824,831</u> |
| % of Ministry of Finance | | | 93.90% |
| Population in 2021 | | Davies 2022 | 982,000 |
| €/citizen in 2021 | | | 114.89 |

– 29 –

| Bremen | Bremen | Bremen | | 2022 |
|---|---|---|---|---|
| **Finances** | 09 Finanzen | p. 76 | 3.557.533.200,41 | 3,557,533,200 |
| of which only for tax administration: | | | | |
| Office of the Senator for Treasury | 0900 Behörde d. Sen. für Finanzen | p. 74 | 62.665.672,04 | 62,665,672 |
| General authorizations for finances and personnel | 0901 Allgemeine Bewilligungen für Finanzen und Personal | p. 74 | 61.351.045,58 | 61,351,046 |
| Landeshauptkasse Bremen | 0910 Landeshauptkasse Bremen | p. 74 | 9.051.494,60 | 9,051,495 |
| Central education, training and further education | 0922 Zentrale Aus-, Fort- und Weiterbildung | p. 74 | 18.517.238,66 | 18,517,239 |
| Administrative School | 0923 Verwaltungsschule | p. 74 | 1.253.454,86 | 1,253,455 |
| Education and Training Center | 0926 Aus- und Fortbildungszentrum | p. 74 | 5.615.298,24 | 5,615,298 |
| College of Public Administration | 0927 Hochschule für Öffentliche Verwaltung | p. 74 | 2.765.513,04 | 2,765,513 |
| IT – Budget | 0950 IT – Budget | p. 74 | 89.062.356,28 | 89,062,356 |
| Bremen-Nord tax office (until April 30, 2017) | 0954 Finanzamt Bremen-Nord (bis 30.04.2017) | p. 74 | 0,00 | 0 |
| Tax Office Bremerhaven | 0955 Finanzamt Bremerhaven | p. 75 | 11.068.398,52 | 11,068,399 |
| Tax Office for External Audit Bremen | 0957 Finanzamt für Außenprüfung Bremen | p. 75 | 8.340.176,18 | 8,340,176 |
| Tax Office Bremen | 0958 Finanzamt Bremen | p. 75 | 16.969.129,75 | 16,969,130 |
| Taxes | 0970 Steuern | p. 75 | 4.323.553,90 | 4,323,554 |
| Only pure tax administrative costs: | | | | 290,983,332 |
| neutral | | | | |
| Central budgeted personnel expenses | 0990 Zentral veranschlagte Personalausgaben | p. 75 | 36.320.860,88 | 36,320,861 |
| General | 0995 Allgemeines | p. 76 | 10.366.535,93 | 10,366,536 |

| | |
|---|---|
| Ratio pure-tax to (pure-tax & pure-non-tax): | 8.29% |
| 'Pure tax admin cost' + 'Ratio pure-tax to (pure-tax & pure-non-tax)' * 'neutral costs': | 294,852,842 |
| % of "Finances" | 8.29% |

more:

| | | | | |
|---|---|---|---|---|
| | 0160 Finanzgericht | p. 43 | 987.618,90 | 987,619 |

| | | | |
|---|---|---|---|
| New total: | | | <u>295,840,461</u> |
| Population in 2021 | | Davies 2022 | 676,000 |
| €/citizen in 2021 | | | 437.63 |
| **Federal and states' governments combined** | All (underlined values) above, totalling: | | **<u>17,667,088,104</u>** |
| Population in 2021 | | Davies 2022 | <u>83,236,000</u> |
| €/citizen in 2021 | | | <u>212.25</u> |